# SYNTHETIC BRAIN IMAGES: BRIDGING THE GAP IN BRAIN MAPPING WITH GENERATIVE ADVERSARIAL MODEL

Drici Mourad[1] and Dr. Kazeem Oluwakemi Oseni[2]

[1]School of Computer Science and Engineering, University of Westminster, London, United Kingdom

[2]School of Computer Science and Technology, University of Bedfordshire, United Kingdom

## ABSTRACT

*Magnetic Resonance Imaging (MRI) is a vital modality for gaining precise anatomical information, and it plays a significant role in medical imaging for diagnosis and therapy planning. Image synthesis problems have seen a revolution in recent years due to the introduction of deep learning techniques, specifically Generative Adversarial Networks (GANs). This work investigates the use of Deep Convolutional Generative Adversarial Networks (DCGAN) for producing high-fidelity and realistic MRI image slices. The suggested approach uses a dataset with a variety of brain MRI scans to train a DCGAN architecture. While the discriminator network discerns between created and real slices, the generator network learns to synthesise realistic MRI image slices. The generator refines its capacity to generate slices that closely mimic real MRI data through an adversarial training approach. The outcomes demonstrate that the DCGAN promise for a range of uses in medical imaging research, since they show that it can effectively produce MRI image slices if we train them for a consequent number of epochs. This work adds to the expanding corpus of research on the application of deep learning techniques for medical image synthesis. The slices that are could be produced possess the capability to enhance datasets, provide data augmentation in the training of deep learning models, as well as a number of functions are made available to make MRI data cleaning easier, and a three ready to use and clean dataset on the major anatomical plans.*

## 1. INTRODUCTION

This Project explore the generation of realistic MRI images slices using Deep Convolutional Generative Adversarial Networks. The proposed method trains a DCGAN architecture using a clean and prepared dataset of sagittal brain MRI scans. The generator network learns to synthesise realistic MRI image, while the discriminator network distinguishes between manufactured and genuine ones. Through an adversarial training strategy, the generator improves its ability to produce slices that closely match real MRI data. This work contributes to the growing body of research on the use of deep learning methods for synthetic medical imaging.

The research objectives proposed are as follows:

International Journal of Managing Information Technology (IJMIT) Vol.16, No.1, February 2024

1- Select and prepare the data with the aim of providing a clean and ready to use dataset containing sliced brain images in the three anatomical directions.
2- Develop and provide python functions that can be turned into a library to make it easier to deal with the Nifty type files and make any future data cleaning and preparation process easier
3- Develop a DCGAN based architecture capable of generating synthetic MRI slice images
4- Evaluate the generated MRI images qualitatively, comparing them against ground truth images.

## 2. LITERATURE SURVEY

### 2.1. MRI

From infancy to adulthood, humans undergo substantial physical and mental changes, and throughout the years, a range of perceptual, cognitive, and motor functions mature. Merely examining behavioural alterations alone is insufficient to gain a deep grasp of such developmental processes, concurrent examination of brain development may provide a more thorough insight. This understanding is largely based on recent developments in non-invasive neuroimaging technologies.

Structural MRI refers to any MRI scan that is used to image the structure and called anatomical scans. fMRI scans are intended to quantify a particular facet of physiological function, usually brain activation and will not be needed in our study, the anatomical structures are described using imaginary planes called anatomical planes, the three most frequently utilised planes: sagittal, coronal, and axial [1]

- The sagittal plane is a vertical plane that runs longitudinally through the body. It separates the left and right halves of the body.
- The coronal plane is a vertical plane that runs longitudinally through the body right angle to the sagittal plane, it creates an anterior (front) and posterior (back) division in the body.
- The axial plane is a horizontal plane. It is parallel to the ground and perpendicular to the coronal and sagittal planes. It separates the body into two sections: the lower, inferior section, and the upper, superior section.

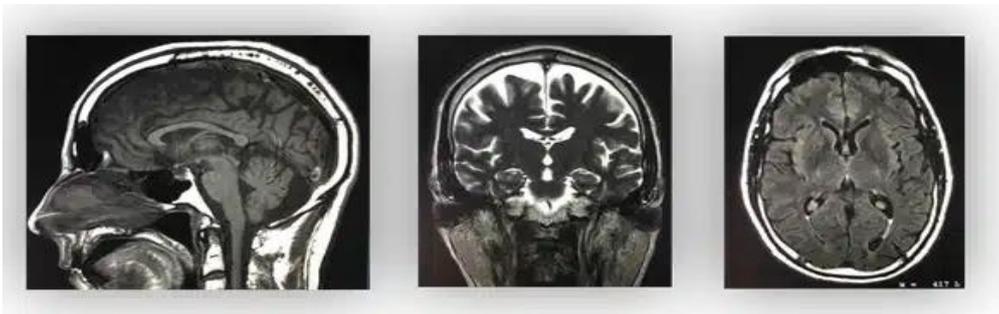

Figure 1. Brain MRI scan, left to right: Sagittal, Coronal, and Axial

### 2.2. Data Preparation in Healthcare

A comprehensive guidance on how to prepare medical images before using AI algorithms has been submitted in the related study [2].

Image de-identification is mostly for patient privacy preserving because consensus consider patient data as highly sensitive resource. The de-identification process consists mostly of two major points, the Meta data which consists of removing all patients identifiers as names and addresses, and removing the voxels data of the images containing facial features like the eyes and the nose, because modern computer vision algorithms could be used to identify these individuals [3]

## 2.3. Generative Adversarial Network

The second part is about what Yann LeCun (VP and Chief AI scientist at Meta) described as "the most interesting idea in the last 10 years in Machine Learning", Goodfellow's Generative Adversarial Network (GAN).

The proposed idea involves using an adversarial process to train two models simultaneously, the Nash equilibrium in game theory serves as the foundation for GAN. It is assumed that there are two neural networks playing the game: a generator and a discriminator. While the discriminator seeks to accurately identify whether the input data is from the generator or the real data, the generator seeks to learn the distribution of the real data to create more accurate generated images. The two neural networks must always better themselves to increase their individual capacities for generation and discrimination aiming to win the game. The goal of the optimisation procedure is to get the two parties to a Nash equilibrium, without using Monte Carlo Approximation nor Markov's chains theory [4].

We will mostly be interested in its DCGAN variant, where deep convolutional layers are incorporated into both the generator and the discriminator which help stabilise the learning process, with the use of batch normalisation to facilitates the training process acceleration and stabilisation [5]

## 2.4. Generative Adversarial Network Variants

Following Goodfellow's GAN idea, also called the Vanilla GAN, other GAN variations have been developed by improving the theoretical part and the model's structure, let's discuss about some of them and show the important differences:

### 2.4.1. DCGAN

Deep convolutional layers are incorporated into both the generator and the discriminator which help stabilise the learning process, batch normalisation is frequently used to facilitates the training process acceleration and stabilisation, and they are very powerful for images generation [5]

### 2.4.2. Wasserstein GAN (WGAN)

When using the gradient descent approach to train GANs, there may be very little overlap between the generated and real data, which results in a constant Jensen-Shannon divergence ofthe objective function causing a vanishing gradient problem, in order to solve this problem, that divergence is replaced with the Earth Mover distance in the Wasserstein GAN for determining the distribution distance between the generated and real data, and to represent the discriminator, a critic function is used to expands upon the Lipschitz constraint. While WGAN is a major step towards robust GAN training, it can still produce poor-quality samples or not converge in some situations [6]

### 2.4.3. Improved WGAN

In certain situations, WGAN produces poor-quality samples or struggles to converge mostly due to the weight clipping to impose a Lipschitz constraint which has the potential to cause abnormal behaviour. An improved approach was suggested to enforce the Lipschitz constraint by penalising the norm of the critic's gradient regarding its input rather than trimming the weights. And this led to a quicker convergence and a greater generation quality [7]

### 2.4.4. Loss-Sensitive GAN (LSGAN)

Another problem with GANs is that, despite their complexity, the discriminator can discern between generated and real samples with unlimited modelling ability, which can lead to over-fitting. The LSGAN was suggested to satisfy the Lipschitz constraint with an objective function to restrict the discriminator's modelling capacity. Both WGAN and LS-GAN enhance the parameter learning and optimisation process. [8]

### 2.4.5. Implementation Review

DCGAN is a very powerful GAN structure when it comes to image related classification and generation because it uses filters that can detect and reproduce various essentials patterns. So, this is probably the way to go. I will aim to follow the same steps as the [5] paper suggest, but if the outcome is not very good and if I don't have any further limitations, and Wasserstein GAN loss function implementation or the improved one could be beneficial for this project.

## 2.5. Generative Adversarial Network in MRI generation

GANs can be trained to generate new samples and mimic the distribution of an initial dataset, which make it a very powerful tool for data augmentation and synthetic image generation, from[9]139 studies out of 789 search results were reviewed. And it showed that the most popular use of GANs is for data augmentation and generation of synthetic brain MRI images. It also has a purpose of brain images classification like brain tumours segmentation, and image transformation where healthy photos were converted to diseased images for example. According to their findings, GANs could improve the way AI techniques are performed on brain MRI,most of these studies were published in 2020 and 2021, and only few between 2016 and 2017(6 for more precision). This is do show the interest in using GANs in neuroimaging and validate that this is a trendy topic.

The Work from [10] introduces a Deep Convolutional Generative Adversarial Network (DCGAN) architecture-based to produces 256 x 256 2D MRI pictures with an axial view of the brain, this was a very interesting article because our objective is to generate 256 x 256 images as well and this is not very conventional for GAN's in general, most of the times it is used on images with a very small resolution.

The purpose of the work made by [11] is to generate MRI scan images by the application of two key generative models, a Vanilla GANs and a DCGANs. This approach yielded good outcomes. As a result, the generated picture validation is done using transfer learning Models as CNN, ResNet152V2, and MobileNetV2. The accuracy score for the output and the efficacy of the classification are assessed using a confusion matrix. For example, the CNN demonstrates how well the model can distinguish between the two brain statuses (tumour and no tumour), which imply that the image generation was a success.

The literature review provided by [12] helped me to comprehend the function of synthetic dummy images generated by GANs. They conducted a thorough search of the Web of Science and Scopus databases to locate pertinent studies over the previous six years.

Composed a review of the systematic literature, Pre-established inclusion and exclusion standards made it easier to narrow down the search results. They identified the different loss functions that are used to process brain MRIs. Selected the appropriate assessment metric for an application and they do consider that GAN-synthesised images will play a major role in the clinical field, the article provides a baseline for this research.

The work provided by [13] is an extensive overview of modern GAN methods and assesses how well they produce artificial medical images to supplement the limited training data available. A variety of medical imaging datasets were used to train six widely used GAN architectures. Significant variations in output fidelity and diversity are revealed by combining a quantitative picture analysis approach with a methodical hyper parameter optimisation strategy. The performance of the downstream segmentation operation offers further domain-specific evaluations of the resulting datasetsutility. According to the study, although some sophisticated GANs can generate medical images that appear realistic, the artificial data constantly performs worse than genuine datasets on specific tasks. The findings warn against using GAN-produced medical images carelessly but also point to ways to create customised GAN solutions for better training.

## 3. DATA SELECTION

The first step is to construct a unified brain middle slice dataset by selecting the right data and making the right alterations, there is no need to do any feature selection as our final dataset will mostly contain brain images slices on the selected anatomical plane only, it is important to be sure that the images represent what we are aiming to generate, with minimum noise and that the overall image quality is good.

OpenNeuro is the principal website where we are getting the different dataset from, Using the search MRI portal is useful because it helps me precise what I need in the dataset:

1. Dataset containing a full brain which can be properly sliced in any anatomical view.
2. Age of Participant contain all ages to have a diversified dataset.
3. Number of participants is set from 100 to 200 to get dataset that went through the exact treatment and curation phases, and this simplify the global data unification.
4. For the Diagnosis, we are choosing Healthy or in Control individuals, because our objective is to generate slices of a healthy brain.
5. We want to generate a human brain, so the category Human in specified.

It is also more beneficial to choose datasets that had a Cross-sectional purpose to avoid data redundancy due to its repeatability overtime in the longitudinal study case, most of the scan acquired used a device at 3T frequency equipped with a 32-channel head coil.

## 4. DATA EXPLORATION

After downloading all the datasets , all non-useful heavy directories containing non needed files were removed and all data undergo a data exploration process to identify all the data alteration needed to clean the datasets to create a unified one, our primary concern are the middle slices, What is considered as the "Middle Slice" is the view we will have if a cutting plane line is put in the middle of the brain, we are supposed to have 3 different middle slice views, one for each anatomical direction.

The most crucial stage in our pipeline is image cleaning and preparation because it sets up the ideal learning environment for our DCGAN , the standard of the input we submit will largely determine the quality of the outcome, the portion of the data that is corrupted or do not represent the feature of a healthy brain were removed, we also were able to generate from every Nifty file multiple central slices in the three anatomical plans, and transforming them into NumPy arrays, as well as resizing every generated image to a consistent resolution of 256x256 pixels and normalizing all the pixel values to fall within the a specific range because this will be the input of our DCGAN. All the generated datasets will be joined into a single one to be transformed into a usable tensor file to create our final data loader.

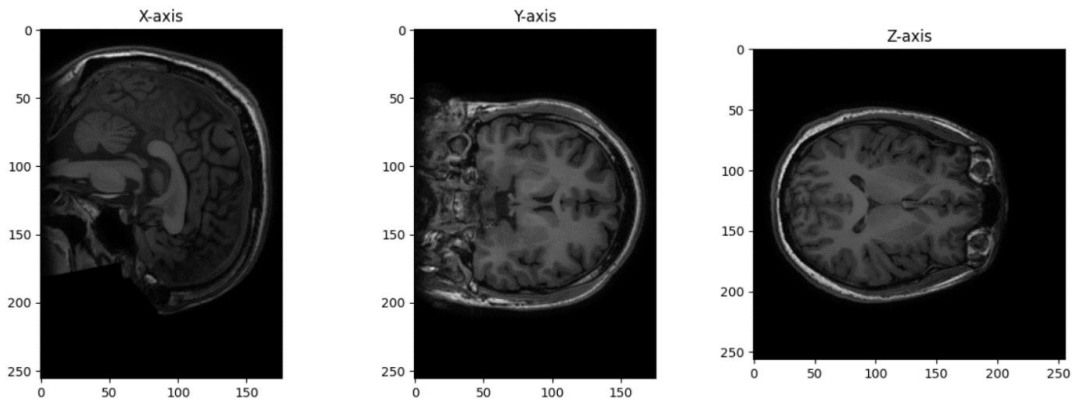

Figure 2. Cerebrovascular dataset middle slices exploration

From the left to the right , we have middle slices corresponding to the sagittal , axial and coronal planes for the cerebrovascular dataset, The coordinate corresponding to each view are x and y = 112 for the sagittal and coronal , and z = 88 for the axial plane, locating these middle slices is important because they are the most representative of what a Brain MRI image is, From each Nifty file , the 15 nearest slices from the middle one will be saved as new images and will be the core inputs of our DCGAN, and as we can see the newly generated pictures will need a 90 degree anti-clockwise rotation. The Dimension of the dataset was able to be retrieved by using a custom function, and we found that it is complex for the sagittal and the coronal {(256, 146), (256, 162), (256, 152), (256, 174), (256, 164), (512, 186), (512, 176), (256, 186), (256, 176), (256, 160), (256, 166), (256, 172), (256, 156)} , custom pixel augmentation and pixel reduction function were applied to homogenise all these pictures into a (256,256) shape, the Axial is {(512, 512), (256, 256)}, so we just need to apply our pixel reduction function to reduce the (512, 512) t o a (256, 256) dimension.

## 5. DATA CLEANING AND PREPARATION

Several custom functions were created in order to make the data cleaning process doable, like reading and plotting the images, middle slices saving and applying images rotations.

- A size verification function to verify the size of all the images in a particular dataset, before and after any dimension alteration to know what type of transformation any image need (pixel reduction or augmentation) and to know if the implementation worked and that the final image size is 256x256.

- Pixel augmentation involves adding empty values to every edge of the image in order to convert a picture into a 256x256 dimension, Padding the image evenly on the top and bottom, as well as on the left and right is mandatory, in order to keep the image central and avoid making any significant changes to it, the noise generated by the padding will not affect the data quality in any way, because no important image features will be disrupted by this manipulation, and we are just adding more black background to achieve our aimed dimension.
- The Pixel Reduction method uses the slicing property on NumPy arrays to remove the null padding after making sure that the cropping will not remove any significant portion of the image.
- In order to streamline and automate the process, we chose to combine all of the data transformation functions into one that could use on every anatomical plan for every dataset, simply entering the appropriate parameters into this function will save the result into a final, homogenised, ready-to-use dataset folder.
- Most of the images in each plan are rotated, either by 180 or 90 degrees in the clockwise and anticlockwise directions, or simply flipped , 4 functions let us get rid of this problematic by putting all the images in the same direction.
- Last custom functions created were about dataset normalization between [-1,1] because not all files had same maximum to be able to automate the normalization and transforming the normalized images into a usable tensor.

Here is an overview of the previous dataset before and after our alterations:

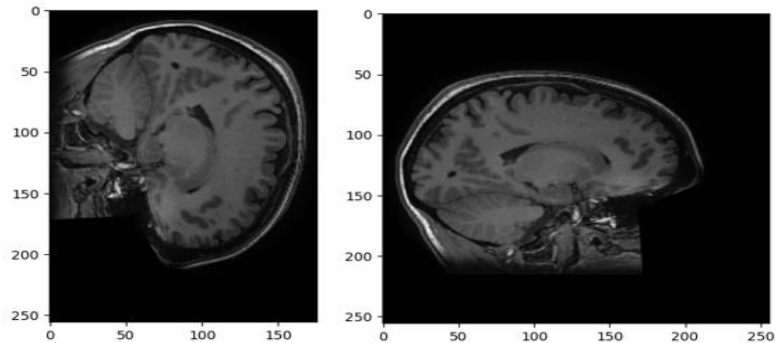

Figure 3. Sagittal files before and after alteration

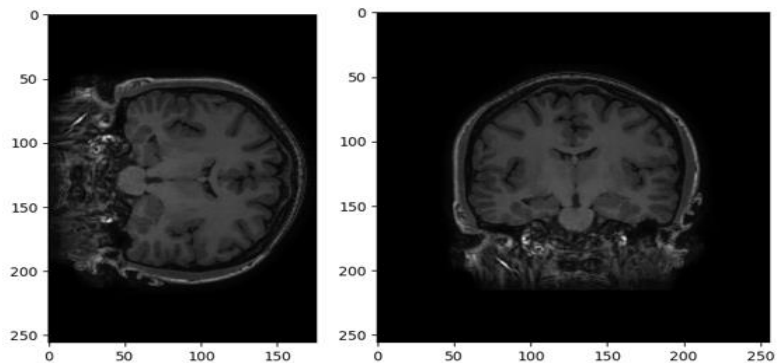

Figure 4. Coronal files before and after alteration

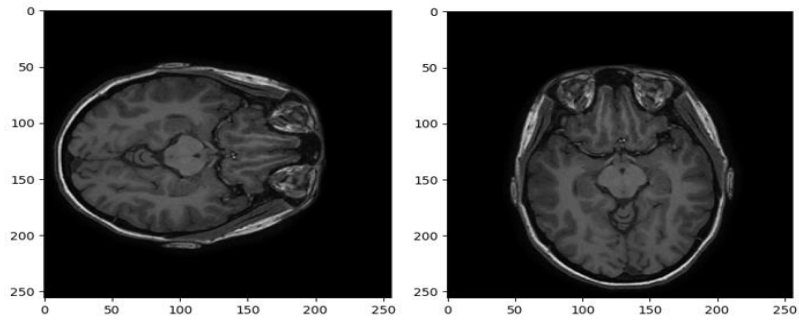

Figure 5. Axial files before and after alteration

## 6. DCGAN STRUCTURE

DCGAN structures are mostly composed by Convolutional neural networks that are capable of handling structured grid data. In tasks such as object detection, picture recognition, and image classification, they have demonstrated remarkable efficacy, due to their strength in feature extraction.

### 6.1. Generator Architecture and Loss Function Implementation

The generator will take as an input a random noise from the normal distribution with a size of 512, the objective is to pass this flattened noise vector, augment it and reshape it using CNN's until we have a 256x256x1 output picture, beside the output layer which will take "Tanh", all the activations functions are "ReLU" as specified in DCGAN paper, the input will pass first in a deconvolution layer with 256 filters which represent the dimensionality of the output space and the windows extraction for pattern recognition, a kernel size of 4 specifies the height and width of the 2D convolution window to finally have a first output of (None, 4, 4, 512), with None the representation of our batch size.

This layer is followed by a convolutional layer this time followed by a Batch normalization to improve the training stability and speed of deep neural networks learning process [9] , we also introduced an Up sampling layers in the end, and these do have the purpose of doubling the dimension of the image, to make it bigger until we achieve our 256x256 goal, we keep the same parameters and only the filter is switched to 128 to have an output or (None, 16, 16, 128) in the next layer then the kernel size is changed from 4 to 3, and this help get more precise windows for the final generations, and we switch our filter from 128 to 64, and divide this value by 2 until we achieve a size of 8. Once the output shape of (None, 256, 256,8) is achieved, we just put it into a last convolutional layer with a filter value of 1, to switch it into a grayscale type output, and define our final output activation function which is "Tanh".

We can test our structure by imputing a random noise vector and see the output without any training.

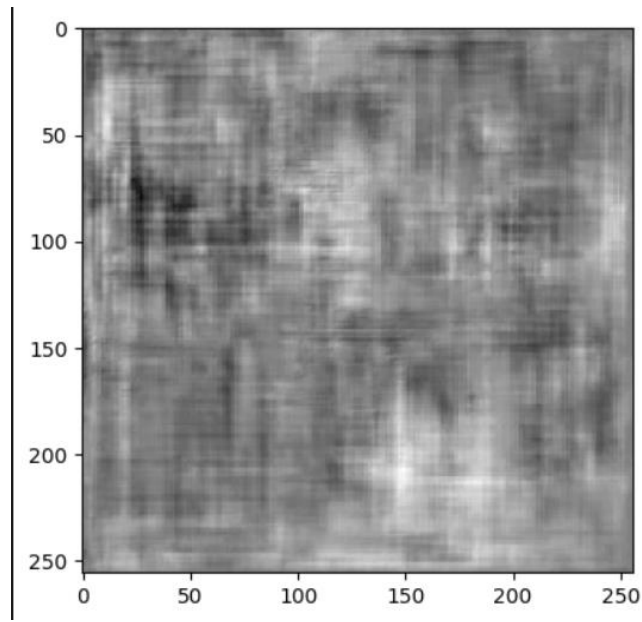

Figure 6. Generated image with no prior training

As we can see the image is a grayscale noise and do have the right (256x256) dimension. When it comes to the loss function, we will use the binary cross-entropy and the Adam optimiser with a learning rate of 0.00001.

## 6.2. Discriminator Architecture and Loss Function Implementation

The first step is adding a Gaussian noise in the first layer of the discriminator, and this is to prevent it from overfitting[14], all activations functions used are "LeakyReLU" as specified by the literature, beside the output layer which takes the "Sigmoid" activation function, to handle the probabilistic behaviour of our mode, 1 for real image, 0 for a fake one, the rest of the discriminator layers are composed of image classifier based on CNN architecture, so a convolutional2d as the second part of the discriminator, witch an uplifted filter at each phase, a constant padding (null) and a kernel size of 3.

Each layer also contain a batch normalization component as suggested by [14] ,a 0.25 dropout is introduced, it's a regularisation technique called to approximates the process of training several neural networks with various designs concurrently [15], and specify that it improve stability and prevent overfitting [16],finally, the AveragePooling2D operation is one of the best down sampling methods, which is the objective of the discriminator, going from a full 256x256 image to a single output node.

Binary cross-entropy and the Adam were used as Loss function and optimiser, In this case we need to calculate two types of loss, one for a real image and one for a fake one, and then sum for out total loss, which will determine how good out discriminator can distinguish between real and fakes ones

International Journal of Managing Information Technology (IJMIT) Vol.16, No.1, February 2024

## 7. RESULTS PRESENTATION

### 7.1. Results Obtained After 50 Epochs

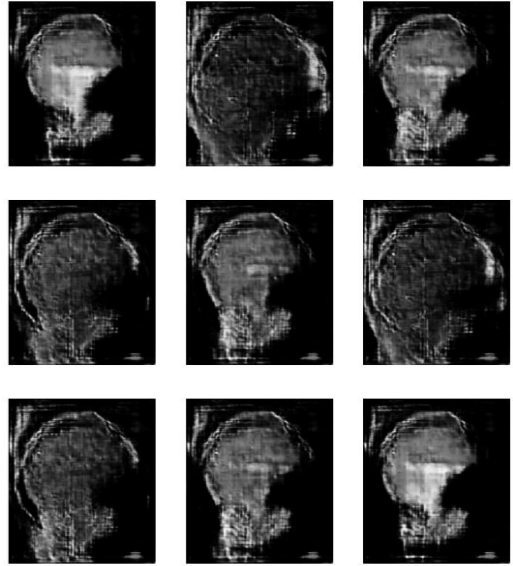

Figure 7. Epoch's 50 output

As we can see the output of this epoch is just trying to poorly mimic the shape of a sagittal plane brain MRI, there is still a lot of noise and none of the image is completed, but this is the expected situation, actually the algorithm is showing good results because it is a very low training process, so being able to recognise some fundamental patterns and generate them from noise is an amazing progress.

### 7.2. Results Obtained After 100 Epochs

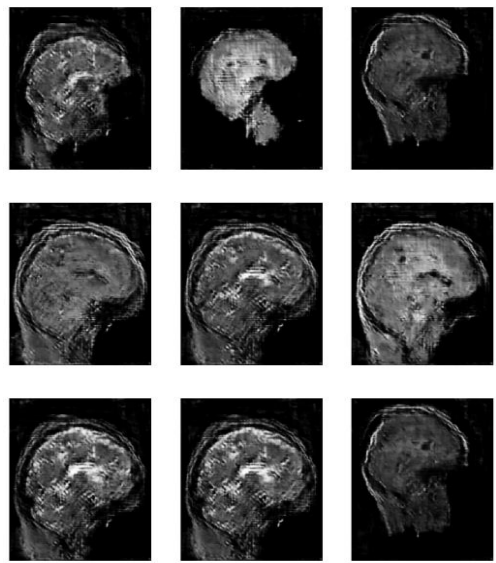

Figure 8. Epoch's 100 output

International Journal of Managing Information Technology (IJMIT) Vol.16, No.1, February 2024

There is still a lot of uncompleted structural pictures, but the shape is preserved, we can also see that the images generated are somehow different, but we can notice that some of the outputs arestill a lot pixelized. Our Dataset is not that diverse to create and expect different type of MRI facets or anything, so it is normal that all the pictures outputted looks alike.

## 7.3. Results Obtained After 200 Epochs

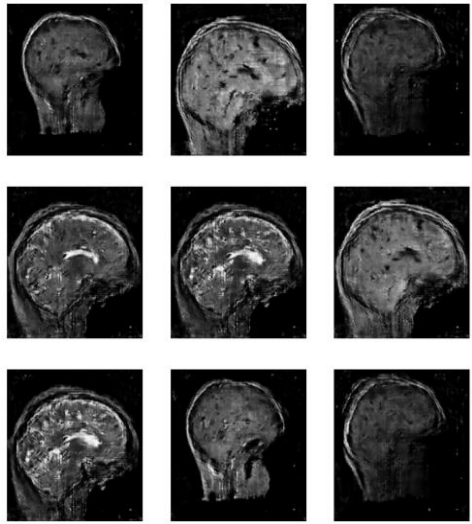

Figure 9. Epoch's 200 output

After 200 epochs, the generator can create a good structural image of the sagittal plane, we can notice that the images output is from different sizes, there is more intention in generating more details in brain tissue, the brain sagittal structure has no secrets anymore to our generator as it is more and more confident to create something coherent to what we are looking for.

## 7.4. Results Obtained After 300 Epochs

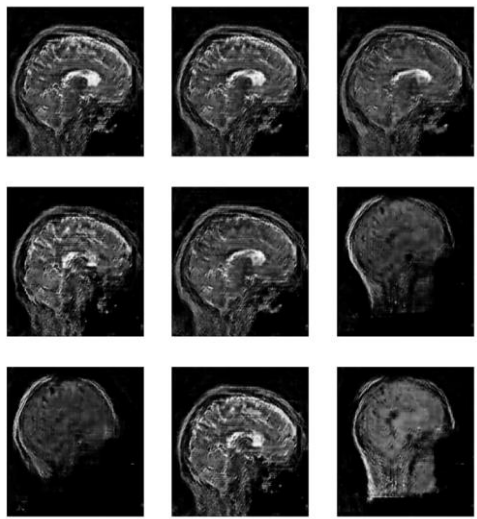

Figure 10. Epoch's 300 output

After 300 epochs, the generator is now confident to create a good structural image of the sagittal plane, there are still not much details on brain tissue unfortunately, but we can see that the algorithm is slightly more focusing on that part of the generation now, we can see that the outside structure is clean and that most of the resonance is in the inside, which will probably result to a generate of more accurate brain tissue details if we kept the training longer.

### 7.5. Qualitative Evaluation Approach

The way chose to evaluate the performance of this DCGAN is qualitative by manually inspecting and judging the output. To distinguish between what is and is not practical, we usually need domain knowledge. It is crucial in our situation to rely on a neuroscientist assistance in evaluating of the output samples. But unfortunately, the training wasn't that intensive and advanced to have plenty of details to rely on and have a more complex interpretation of the results, but we are still able to compare the output with the input and conclude that the algorithm is doing well. When it comes to the Quantitative method usually used in GAN, it is about the use of the Inception score, or one of the derived functions. The issue with the inception score is that it relies on two parameters, the first one is the quality of the generated output, which is not that great in that case due to the low number of epochs in the training, and the second one is the image diversity [17]. Our Objective was to generate MRI images, and this could never have a diverse approach in the generation, it not the same purpose as generating cats or dogs in multiple environments, it will always be the same brain structures with just different scales or small details. The best way to evaluate medical image generation is using the qualitative way with a professional expertise.

## 8. CONCLUSION

The scripts and custom functions created will assist anyone working with MRI datasets, help in the visualization of middle slices of any Nifty file in the three anatomical directions, take snapshots of entire heavy brain MRI images into 2D slices in any desired anatomical plan, as well as verify the size of each image created to have a better idea of the appropriate transformations and apply them, three clean and usable datasets with more than 18000 central slices in each of the coronal, sagittal and axial plane is another beneficial result of this study.This implementation is a ready to use DCGAN structure that can be used with any 256x256 grayscale image input. Our goal was accomplished once we were able to create a synthetic structural MRI image from full structural MRI nifty raw data on the chosen plane (in our case the sagittal one). The structural aspect completely corresponds to what we should have, with the exception of a lack ofdetails due to the low number of epochs our model was trained on.

## 9. OPTIMISATION AND FUTURE WORK

The first Optimisation that can be done on the work is train it more, for at least 1200 epochs and evaluate the output. It would be beneficial to take the script as it is and just try it with the axial and coronal dataset as input to see if the results are good and comparable to the one, we can get with the sagittal one. Multiple learning rate parameters can be tried to see if there is a huge difference between the obtained outputs, because the learning rate of the Adam optimiser is the most important hyper parameter in this case, another simple idea would be to implement a learning rate that becomes smaller over time so that the model converge more effectively.
Also, another optimisation could be [11]:

- Use of the "LeakyReLU" instead of "ReLU" for the generator structure.
- Noise sampling from a Gaussian distribution rather than a normal one [9]
- Adding noise to the first layer of the generator as we did for the discriminator.

This work was mostly based on the DCGAN paperwork, but the implementation of a slightly different structure and loss functions can lead to far better results, the structure and recommendation of the Wasserstein Gan (WGAN) paper [7], the improved version [8], and the LSGAN can all be used with the same datasets and compare the performances.

The first suggested dataset optimisation would be to apply the tensorizing script on the entire dataset rather than just taking 10 000 pictures to have a lot more data in each epoch, more data means more training and better generation.

The second suggestion would be to use some data augmentation, and slightly include some very small rotations, zoom of unzoom some pictures, as well as playing with the contrast to alter minimally the images.